# A criterion for "easiness" of certain SAT-problems

Bernd R. Schuh

Dr. Bernd Schuh, D-50968 Köln, Germany; bernd.schuh@netcologne.de

keywords: complexity, satisfiability, propositional logic, P, NP, 1-in-3SAT, easy/hard instance

**Abstract.**

A generalized 1-in-3SAT problem is defined and found to be in complexity class P when restricted to a certain subset of CNF expressions. In particular, 1-in-$k$SAT with no restrictions on the number of literals per clause can be decided in polynomial time when restricted to exact READ-3 formulas with equal number of clauses (m) and variables (n), and no pure literals. Also individual instances can be checked for "easiness" with respect to a given SAT problem. By identifying whole classes of formulas as being solvable efficiently the approach might be of interest also in the complementary search for "hard" instances.

**Introduction.**

Many problems in propositional logic are varieties of the decision problem $F \in$ SAT ? and are in complexity class NP. Examples are 1-in-3SAT which is the problem of deciding whether for a given 3-CNF formula there exists an assignment which evaluates exactly one literal per clause to *true*, or NOT-ALL-EQUAL-3SAT which asks for an assignment with at least one true and one false literal per clause. Others, like e.g. HORN-SAT or 2-SAT, are known to be decidable in linear time and thus belong to complexity class P. For these and other examples see e.g. [1].

An extension of some of these NP problems to a more general requirement on the number of true literals and to instances where the number of literals in each clause is not restricted to exactly 3 will in general enhance the complexity of the problem. One can identify restricted CNF expressions, however, for which the problems lie in complexity class P.

The basic idea is as follows. Given a SAT problem which typically asks the question "Does a truth assignment exist with property X?". If you manage to find in time polynomial a limited set of assignments which are the only ones to fulfill a necessary condition of property X then they form the only certificates which need to be tested on property X, and the whole process is done efficiently. In the following I will formulate conditions which allow to determine such a limited set for certain SAT problems. The criterion can be evaluated in time polynomial and be used to determine a given instance as "easy" with respect to the problem considered.

**Notation.**

A Boolean formula in *Conjunctive normal form* CNF by definition is a conjunction of clauses, where each clause is a disjunction of literals. A literal is an occurrence of a Boolean variable (*atom, basic variable*) or its inverse/negative/negated. The Table lists some parameters by which a general CNF formula *F* may be characterized, though not completely.

| | |
|---|---|
| $m$ | number of clauses |
| $n$ | number of variables |
| $k_j$ | number of literals in clause $C_j$ |
| $p_s = p_{s+} + p_{s-}$ | occurrence of atom $a_s$ (also called "degree", "frequency" or "appearance") = sum of negative (negated) and positive (unnegated) literals of variety *s* |
| $m_\alpha$ | number of clauses $C_j$, for which $k_j=\alpha$ |
| $n_\beta$ | number of atoms $a_s$, for which $p_s=\beta$ |
| $N$ | total number of literals |
| $N_+$ | number of positive literals |
| $N_-$ | number of negative literals |

The following relations hold for CNF expressions in terms of the above quantities.

$$m = \sum_\alpha m_\alpha \; ; \; n = \sum_\beta n_\beta$$

$$N = N_+ + N_- = \sum_{s=1}^{n} p_s = \sum_{j=1}^{m} k_j = \sum_\alpha \alpha m_\alpha = \sum_\beta \beta n_\beta$$

In the following we use the term $\{\leq k, \leq p\}^{(m,n)} - SAT$ *instance* or $\{\leq k, \leq p\}^{(m,n)} - CNF$ for a Boolean CNF expression with *m* clauses, *n* variables and no more than *k* literals per clause and no more than *p* occurrences per variable. By dropping the $\leq$ prefix we indicate that the expression has exactly *k* literals /*p* occurrences per clause/variable. Instances <u>without pure literals</u>, i.e. $p_{s+} p_{s-} \neq 0$ for all *s* $\in N_n$ will be called *completely mixed*. Note that $\{r, \leq r\}^{(m,m)} -$ or $\{\leq r, r\}^{(m,m)}$ - CNF automatically are exact expressions, i.e. $\{r, r\}^{(m,m)}$ - CNF, due to m=n and the relations $\sum_\alpha \alpha m_\alpha = \sum_\beta \beta n_\beta$.

The following commonly used terms are special cases:

A *3SAT instance* is a Boolean expression in CNF with $k_j$=3 for all clauses, $j = 1, 2, ..., m$, i.e. a $\{3, \leq p\}^{(m,n)} - SAT$ *instance*.

A *READ-3 SAT instance* is a Boolean CNF expression in which all variables have degree 3 or less, i.e. a $\{\leq k, \leq 3\}^{(m,n)} - SAT$ *instance.*

Any $\{\leq k, \leq p\}^{(m,n)} - SAT$ instance can be transformed to a $\{3, \leq p\}^{(m',n')} - SAT$ instance without loss of satisfiability and in polynomial time. Likewise it is possible to transform any $\{\leq k, \leq p\}^{(m,n)} - SAT$ instance to an exact READ-3 CNF, i.e. a $\{\leq k, 3\}^{(m',n')} - SAT$ instance. Combining both reductions leads to a $\{\leq 3, 3\}^{(m'',n'')} - SAT$ or a $\{3, \leq 4\}^{(m'',n'')} - SAT$ instance in the best case. No way from an arbitrary CNF formula leads to a CNF with <u>exactly</u> 3 literals per clause <u>and</u> no more than 3 occurrences of each variable. Tovey noticed that in this sense the $\{3, \leq 4\}^{(m,n)} - SAT$ problem is the "smallest" NP-complete satisfiability problem. In fact, any $\{r, \leq r\}^{(m,n)} - SAT$ instance is satisfiable and thus trivial in a way [2]. The proof uses Hall's theorem [3].

**The *PART-SAT* problem.**

We now define a class of satisfiability problems by

<u>Definition *PART-SAT*</u>:

Let $F \in \{\leq k, \leq p\}^{(m,n)}$ - CNF, and let $\{\mu_i \in \{0,.1,..,m\} | i \in \{0,1,...,k\}\}$ be a partition of *m*, i.e. $m = \mu_0 + \mu_1 + ... + \mu_k$. Does a truth assignment exist such that $\mu_\alpha$ many clauses contain exactly $\alpha$ true literals each?

To relate the problem to the specific partition we will also use the notation $\{\mu_\alpha\} - SAT$. As an example set $\mu_\alpha = 0$ for all $\alpha$ except $\alpha = 1$, i.e. $\mu_1 = m$, and restrict *F* to $\{3, \leq p\}^{(m,n)}$ - CNF. Then $\{0, m, 0, 0\} - SAT$ coincides with 1-in-3SAT.

As a further example let again be $F \in \{3, \leq p\}^{(m,n)}$ and set $\mu_\alpha = 0$ for all $\alpha$ except $\mu_1$ and $\mu_2$. Then deciding $\{0, \mu_1, \mu_2, 0\} - SAT$ for <u>all</u> pairs $\mu_1 + \mu_2 = m$ is equivalent to deciding NOT-ALL-EQUAL-3SAT. One can also use *PART-SAT* to investigate the question whether certain CNF expressions have assignments which leave a given number of clauses unsatisfied, $\mu_0 \neq 0$.

**The criterion.**

The central criterion for identifying p. t. SAT problems is the following

<u>Theorem</u>

If $\{\mu_\alpha\} - SAT$ is restricted to one of two subsets of CNF expressions, either

$$\left\{ F \in \{\leq k, \leq p\}^{(m,n)} \left| \sum_{\alpha=1}^{k} \alpha \mu_\alpha = \sum_s \min(p_{s+}, p_{s-}) \right. \right\} \text{ or } \left\{ F \in \{\leq k, \leq p\}^{(m,n)} \left| \sum_{\alpha=1}^{k} \alpha \mu_\alpha = \sum_s \max(p_{s+}, p_{s-}) \right. \right\}, \text{ it}$$

is decidable in time polynomial times $2^{n_=}$, where $n_=$ is the number of variables with $p_{s+} = p_{s-}$.

As a <u>corollary</u> to the theorem we can state, that

$\{\mu_\alpha\}$ - *SAT* is decidable in time polynomial, if instances are restricted to formulas for which $\sum_\alpha \alpha \mu_\alpha$ equals either the minimum or the maximum number of literals which can be assigned *true*, and for which no variable occurs in equal numbers of positive and negative literals.

**Sum satisfiability $\sigma$ and proof of theorem.**

For any CNF instance *F* with *m* clauses, *n* atoms and literals $l_{js}$, and any truth assignment $T_x : \{\text{atom}_1, \text{atom}_2, ..., \text{atom}_n\} \to \{0,1\}$, numbered by $x \in \{+1, -1\}^n$ as defined in [4], we define the *sum satisfiability* as the total of true literals under assignment $T_x$:

$$\sigma_F(x) = \sum_{j=1}^{m} \sum_{s=1}^{n} T_x(l_{js}) .$$

As a double sum it can be evaluated either by summing over clauses or over variables first:

$$\sigma_F(x) = \sum_{var iables\, s} \sigma_s(x) = \sum_{clauses\, j} \sigma_j(x)$$

in an obvious notation. As a side remark we state, that for an exact READ-3 CNF, $F \in \{\leq k, 3\}^{(m,n)}$, $\sigma_F$ is a particularly simple quantity, because in this case the characteristic function

$$\chi_\sigma(\alpha) := 2^{-n} \sum_{\{x_s = \pm 1\}} e^{\alpha \sigma_F(x)} = 2^{-n} \prod_s (e^{\alpha p_{s+}} + e^{\alpha p_{s-}})$$

simplifies to

$$\chi_\sigma(\alpha) = e^{\alpha n}\{(e^\alpha+1)/2\}^n (2\cosh\alpha - 1)^{n_p}$$

where $n_p$ is the number of pure variables.

Thus for completely mixed exact READ-3 formulas ($n_p = 0$) $\sigma_F - n$ follows a binomial distribution, and consequently the number of assignments $T_x$ under which $\sigma_F(x)$ evaluates to $n+k$ is $\binom{n}{k}$.

We now return to the general case. To prove the theorem consider the circumstances of $\{\mu_\alpha\}$-$SAT$. If there is an assignment $x_0$ with the desired property then it must belong to a set of assignments which fulfill $\sigma_F(x) = \sum_\alpha \alpha \mu_\alpha$ according to the definition of the sum satisfiability. Quite generally $\sigma$ has a minimum and maximum value with respect to all assignments:

$$\sigma_{F\min} := \min_x \sigma_F(x) = \sum_s \min(p_{s+}, p_{s-})$$

$$\sigma_{F\max} := \max_x \sigma_F(x) = \sum_s \max(p_{s+}, p_{s-})$$

The minimum and maximum states are degenerate if there are variables with $p_{s+} = p_{s-}$, the degeneracy being $2^{n_=}$, if $n_=$ denotes the number of variables with $p_{s+} = p_{s-}$. If $p_{s+} \neq p_{s-}$ for all variables then there is exactly one assignment $x_{\min/\max}$ for which $\sigma_F(x_{\min/\max}) = \sigma_{F\min/\max}$ holds, namely $x_{\min/\max,s} = (-/+)\operatorname{sgn}(p_{s+} - p_{s-})$, respectively. If there were more than one such assignment it necessarily would lead to a reduction of $\sigma$ (in case of maximum) or an enhancement (in case of minimum).

Therefore, if $F$ is restricted to the subset $\sum_\alpha \alpha\mu_\alpha = \sum_s \min(p_{s+}, p_{s-}) = \sigma_{F\min}$, then there is just one assignment with this property, except for the $p_{s+} = p_{s-}$ degeneracy which leads to a factor $2^{n_=}$ for the allowed assignments. Determining $\sigma_{\min/\max}$ is a simple counting procedure, working through the $n$ variables. Thus the problem can be decided in the stated number of steps.

The same line of argument works for $F \in \{\leq k, \leq p\}^{(m,n)}$ restricted to $\sum_\alpha \alpha\mu_\alpha = \sum_s \max(p_{s+}, p_{s-})$, of course .

**Illustrative examples.**

The 1-in-3SAT problem was proved to be NP-complete by Schaefer as a special case of Schaefer's dichotomy theorem [5]. Similarly we can argue that the 1-in-$k$SAT problem, i.e. the same problem without restrictions on the number of literals per clause, is NP-hard, as well. With the help of the theorem it is possible to identify a subclass of CNF instances for which the problem can be decided in polynomial time. For 1-in-3SAT, or more generally 1-in-$k$SAT $\sum_\alpha \alpha \mu_\alpha = m$ must be fulfilled, because exactly one true literal per clause is required. Thus, according to the theorem, for $F \in \{\leq k, \leq p\}^{(m,n)}$ with either $m = \sum_s \min(p_{s+}, p_{s-})$ or $m = \sum_s \max(p_{s+}, p_{s-})$ and $p_{s+} \neq p_{s-}$ for all $s$ 1-in-$k$SAT is in complexity class P.

Take as an illustration the formula $(a,c)(a,b,\bar{c},e)(\bar{a},\bar{b},e)(\bar{b},\bar{d},e)(\bar{b},d,\bar{e})(b,d)$. Determining the minimum of the sum satisfiability is a simple counting process. To make this process more clearly arranged we use a notation in terms of the adjacency matrix scheme, see [4] for details. In this notation rows represent clauses and columns variables. A cross × (not to be confused with the assignment index $x$!) at position ($j$,$s$) stands for a positive literal of variety $s$ in clause $j$, $\bar{\times}$ for a negative. 0 at position ($j$,$s$) in the matrix scheme indicates that variable $s$ does not appear in clause $j$. The aforementioned example then reads

$$\begin{array}{ccccc} \times & 0 & \times & 0 & 0 \\ \times & \times & \bar{\times} & 0 & \times \\ \bar{\times} & \bar{\times} & 0 & 0 & \times \\ 0 & \bar{\times} & 0 & \bar{\times} & \times \\ 0 & \bar{\times} & 0 & \times & \bar{\times} \\ 0 & \times & 0 & \times & 0 \end{array}$$

In this representation it is immediately clear that variable $c$ has one positive and one negative literal. Thus there are two assignments which minimize $F$, namely (-1,1,1,-1,-1) and (-1,1,-1,-1,-1). Each must be checked against each clause to determine whether it leads to exactly 1 true literal. Only the first assignment passes this test, the second conflicts already with the first clause.

One may restrict the allowed expressions further to completely mixed exact READ-3 formulas. This way one gets rid of "degenerate" variables with $p_{s+} = p_{s-}$. Then $\sigma_{F\min} = n$ and the only non-trivial candidates for 1-in-$k$SAT expressions which are in P are instances with $m = n_3 = n$. We call such expressions *square* for obvious reasons. The following instances are illustrations of such square CNF.

$$F_1 = \begin{array}{ccccc} \times & 0 & \times & 0 & 0 \\ \bar{\times} & 0 & \bar{\times} & \times & \times \\ \bar{\times} & \bar{\times} & \bar{\times} & \times & 0 \\ 0 & \bar{\times} & 0 & \bar{\times} & \bar{\times} \\ 0 & \times & 0 & 0 & \times \end{array}$$

$$F_2 = \begin{array}{cccccccc} \times & 0 & 0 & \bar{\times} & \times & 0 & 0 & 0 \\ \times & 0 & 0 & 0 & \times & 0 & 0 & \times \\ \bar{\times} & \times & 0 & 0 & \bar{\times} & 0 & 0 & 0 \\ 0 & \times & 0 & 0 & 0 & \times & 0 & \bar{\times} \\ 0 & \bar{\times} & \times & 0 & 0 & \times & 0 & 0 \\ 0 & 0 & \times & 0 & 0 & \bar{\times} & \times & 0 \\ 0 & 0 & \bar{\times} & \times & 0 & 0 & \times & 0 \\ 0 & 0 & 0 & \times & 0 & 0 & \bar{\times} & \times \end{array}$$

$$F_3 = \begin{array}{cccccccc} \times & 0 & \bar{\times} & 0 & 0 & \times & 0 & 0 \\ \times & 0 & 0 & \times & 0 & \bar{\times} & 0 & 0 \\ \bar{\times} & 0 & 0 & \times & 0 & 0 & \times & 0 \\ 0 & \times & 0 & \bar{\times} & 0 & 0 & \times & 0 \\ 0 & \times & 0 & 0 & \bar{\times} & 0 & \bar{\times} & 0 \\ 0 & \bar{\times} & 0 & 0 & \times & 0 & 0 & \times \\ 0 & 0 & \times & 0 & \bar{\times} & 0 & 0 & \times \\ 0 & 0 & \times & 0 & 0 & \times & 0 & \bar{\times} \end{array}$$

All three instances are satisfiable. But not all are 1-in-3SAT expressions. The only assignment which minimizes $\sigma_{F_1}$ is x=(-1,1,1,-1,-1). But it does not satisfy clause 2. Thus there is no assignment that gives one true literal per clause. Since $\sigma_{F_1 \max} = 10 = 2m$ also $\{0,0,m,0\}$ - SAT can be checked with just one assignment, namely (1,-1,-1,1,1). It fails to achieve 2 true literals in the first clause. For $F_2$, $x_{\min} = (-1,-1,-1,-1-1,-1,-1,-1)$ which violates clause 2. Whereas the same $x_{\min}$ obviously leads to exactly one true assignment in each clause for instance $F_3$. Also the answer to the $\{0,0,m,0\}$ - SAT problem is positive now. Again, $\sigma_{F_3 \max} = \sum_\alpha \alpha \mu_\alpha = 2m$ and $x_{\max} = (1,1,...,1)$ is the only assignment to be checked. A look at the matrix scheme reveals that setting all assignments to *true* indeed leads to the desired result.

As a further illustration consider the *PART-SAT* problems $\{0,m,0,...,0\} - SAT$ and $\{0,0,m,...,0\} - SAT$ restricted to $F \in \{3, \leq p\}^{(m,n)}$, and write $\sigma_{F \min/\max}$ as follows

$$\sigma_{F \min} = \frac{1}{2}(N - \sum_s |p_{s+} - p_{s-}|)$$

$$\sigma_{F \max} = \frac{1}{2}(N + \sum_s |p_{s+} - p_{s-}|) \ .$$

For exact 3-CNF $N = 3m$ holds and both PART-SAT problems are efficiently solvable for all *F* which fulfill $m = \sum_s |p_{s+} - p_{s-}|$ and have no degenerate variables. Obviously all square completely mixed $F \in \{3,3\}^{(m,m)}$ as discussed before belong to this class (see formulas $F_2$ and $F_3$ ). A less symmetric example would be

$$\begin{matrix} x & \bar{x} & 0 & \bar{x} \\ x & x & 0 & \bar{x} \\ 0 & \bar{x} & x & x \\ \bar{x} & \bar{0} & \bar{x} & x \\ 0 & \bar{x} & x & \bar{x} \end{matrix}$$

The assignments to be checked are (-1,1,-1,1) for the 1-in-3SAT problem, and (1,-1,1,-1) for the 2-in 3-SAT problem. Both fail to meet the requirement.

An example where the criterion does not identify any easy instance at all is the 2/2/4-SAT problem described in [1]. One searches for assignments which have exactly two true literals in each clause of a square $\{4, \leq 4\}^{(m,m)}$ - CNF, where variables with occurrence 4 have two positive and two negative literals. According to [1] the problem is NP-complete. The following two instances serve as an illustration for *m*=5 and *m*=6:

$$\begin{array}{ccccc} \times & 0 & \times & \times & \bar{\times} \\ \times & \times & \times & 0 & \bar{\times} \\ \bar{\times} & \bar{\times} & \bar{\times} & \times & 0 \\ \bar{\times} & \bar{\times} & 0 & \times & \times \\ 0 & \bar{\times} & \times & \times & \times \end{array} \qquad \begin{array}{cccccc} \bar{\times} & 0 & \times & \bar{\times} & 0 & \times \\ \times & 0 & \bar{\times} & \times & 0 & \bar{\times} \\ \times & \bar{\times} & 0 & \times & \bar{\times} & 0 \\ \bar{\times} & \times & 0 & \bar{\times} & \times & 0 \\ 0 & \times & \bar{\times} & 0 & \times & \bar{\times} \\ 0 & \bar{\times} & \times & 0 & \bar{\times} & \times \end{array}$$

This is a case where minimum and maximum value of the sum satisfiability coincide: $\sigma_{F\min} = 2n = 2m = \sigma_{F\max}$. On the other hand we are dealing with the *PART-SAT* problem $\{0,0,2,0,0\} - SAT$ and two true literals per clause means $\sum_\alpha \alpha \mu_\alpha = 2m$. Thus the assumptions of the theorem are fulfilled. But the degeneracy is maximal now, $n_= = n = m$. So all $2^n$ assignments are candidates to be tested in principle. In fact, for the *m*=5 instance there are two assignments with the required property, namely (-1,1,1,1,1)=(false, true, true, true, true) and its negation.

Also NOT-ALL-Equal-SAT (NAE-SAT) can not be simplified with the help of the theorem. NAE-SAT asks for assignments which lead to at least one true and one false literal in each clause. If the set of allowed instances is restricted to $F \in \{3,3\}^{(m,n)}$ - CNF, the searched for assignment must deliver either one or two true literals per clause, and thus NAE-SAT is equivalent to deciding $\{0,\mu,m-\mu,0\} - SAT$ for all $\mu = 0,1,...,m$. This in principle is a O($2^m$) task.

This problem can also be put the following way. Since for exact 3-CNF the number of true literals in clause *j* can only take values 1 or 2 for NAE-assignments, which is equivalent to $\sigma_j(3-\sigma_j) = 2$, the equation

$$3\sigma(x) - \sum_j \sigma_j(x)^2 = 2m$$

is a necessary condition for assignments *x* which solve the NAE-3SAT problem. In terms of adjacency matrix elements the condition may be written:

$$m + \sum_{\langle s,s'\rangle} \mu_{ss'} x_s x_{s'} = 0 \quad with \quad \mu_{ss'} \equiv \sum_j f_{js} f_{js'}$$

Any NAE assignment is to be found among the solutions of this equation. Although the $\mu_{ss'}$ are easily calculated in p.t. for any given 3-CNF *F*, there is in general no efficient way to determine the allowed *x*. Note that the similar equation which determines the satisfying assignments of the regular 3SAT problem contains additional terms, linear and trilinear in *x*, [4]. Though 3SAT and NAE-SAT belong to the same complexity class NP, one is tempted to say that NAE-3SAT - although it imposes stronger conditions than 3SAT - is somewhat "easier" than 3SAT, since it lacks the trilinear terms.

**Conclusion.**

I have derived criteria for Boolean CNF formulas to be "easy" instances for a class of SAT-problems, termed *PART-SAT*. *PART-SAT* asks for assignments which generate exactly $\alpha$ true literals in $\mu_\alpha$ clauses, and $\sum_\alpha \mu_\alpha = m$, the total number of clauses. The criterion states that an instance *F* is decidable in time polynomial (times a "degeneracy factor", if there are variables with $p_{s+} = p_{s-}$) – i.e. "easy" with respect to the problem posed – provided $\sum \alpha \mu_\alpha$ equals either the minimum or maximum value of the total of true literals, $\sigma_{F\min/\max}$. This latter quantity can be determined in linear time due to the additivity of $\sigma$ in both clauses and variables. In general, it is difficult to use this selection criterion to single out a simply definable class of expressions as *PART-SAT*-"easy", i.e. as a candidate for P-complexity. In case of 1-in-3SAT or more generally *l*-in-*k*SAT with $l \leq k$ which are special cases of *PART-SAT*, such a simple class could be identified, namely the class of square, completely mixed READ-3 formulas . Nevertheless, one can always check individual instances on "easiness" with respect to a given *PART-SAT* problem. The hope is to ease the search for hard instances via this complementary tool, too.